\documentclass[useAMS,usenatbib]{mn2e}
\usepackage{graphicx}
\usepackage{epsf}
\usepackage{amsmath}
\usepackage{subfigure}

\newcommand{\Mpc}{{\rm Mpc}}

\newcommand{\hmpc}{h^{-1}{\rm Mpc}}

\newcommand{\kms}{\;{\rm km}\,{\rm s}^{-1}}

\newcommand{\JM}{J_{\mathrm{M}}}
\newcommand{\JV}{J_{\mathrm{V}}}

\newcommand{\aap}{A\&A}

\newcommand{\apj}{ApJ}
\newcommand{\apjl}{ApJL}
\newcommand{\aj}{AJ}
\newcommand{\araa}{ARA\&A}
\newcommand{\mnras}{MNRAS}

\newcommand{\nat}{Nature}

\newcommand{\hi}{\mathrm{H\,I}}
\newcommand{\hei}{\mathrm{He\,I}}
\newcommand{\heii}{\mathrm{He\,II}}
\newcommand{\heiii}{\mathrm{He\,III}}
\newcommand{\oi}{\mathrm{O\,I}}

\newcommand{\cii}{\mathrm{C\,II}}

\newcommand{\civ}{\mathrm{C\,IV}}
\newcommand{\siiv}{\mathrm{Si\,IV}}

\newcommand{\fesc}{f_{\mathrm{esc}}}

\begin{document}

\title[The $z\sim6$ UVB]{The Soft, Fluctuating UVB at $z\sim6$ as Traced by $\civ$, $\siiv$, and $\cii$}

\author[K.\ Finlator et al.]{
\parbox[t]{\textwidth}{\vspace{-1cm}
Kristian Finlator$^{1,8,10}$, 
B.\ D.\ Oppenheimer$^{4}$,
Romeel Dav\'e$^{5,6,7}$,
E.\ Zackrisson$^3$,
Robert Thompson$^{9}$,
Shuiyao Huang$^{2}$
}
\\\\$^1$ New Mexico State University
\\$^2$ University of Massachussetts, Amherst, MA, USA
\\$^3$ Department of Physics and Astronomy, Uppsala University, 751 20 Uppsala, Sweden
\\$^4$ CASA, Department of Astrophysical and Planetary Sciences, University of Colorado, 389-UCB, Boulder, CO 80309, USA
\\$^5$ University of the Western Cape, Bellville, Cape Town 7535, South Africa
\\$^6$ South African Astronomical Observatories, Observatory, Cape Town 7525, South Africa
\\$^7$ African Institute for Mathematical Sciences, Muizenberg, Cape Town 7545, South Africa
\\$^8$ Dark Cosmology Centre, Niels Bohr Institute, University of Copenhagen, Copenhagen, Denmark
\\$^9$ NCSA, University of Illinois, Urbana-Champaign, IL 61820
\\$^{10}$ finlator@nmsu.edu
\author[The $z\sim6$ UVB]{
K.\ Finlator,
B.\ D.\ Oppenheimer,
R.\ Dav\'e,
E.\ Zackrisson,
R.\ Thompson,
\& S.\ Huang,
}
}

\maketitle

\begin{abstract}
The sources that drove cosmological reionization left clues regarding their identity in the slope and inhomogeneity of the ultraviolet ionizing background (UVB): Bright quasars (QSOs) generate a hard UVB with predominantly large-scale fluctuations while Population II stars generate a softer one with smaller-scale fluctuations.  Metal absorbers probe the UVB's slope because different ions are sensitive to different energies.  Likewise, they probe spatial fluctuations because they originate in regions where a galaxy-driven UVB is harder and more intense.  We take a first step towards studying the reionization-epoch UVB's slope and inhomogeneity by comparing observations of 12 metal absorbers at $z~\sim6$ versus predictions from a cosmological hydrodynamic simulation using three different UVBs: a soft, spatially-inhomogeneous ``galaxies+QSOs" UVB; a homogeneous ``galaxies+QSOs" UVB~\citep{haa12}, and a ``QSOs-only" model.  All UVBs reproduce the observed column density distributions of $\cii$, $\siiv$, and $\civ$ reasonably well although high-column, high-ionization absorbers are underproduced, reflecting numerical limitations.  With upper limits treated as detections, only a soft, fluctuating UVB reproduces both the observed $\siiv/\civ$ and $\cii/\civ$ distributions.  The QSOs-only UVB overpredicts both $\civ/\cii$ and $\civ/\siiv$, indicating that it is too hard.  The~\citet{haa12} UVB underpredicts $\civ/\siiv$, suggesting that it lacks amplifications near galaxies.  Hence current observations prefer a soft, fluctuating UVB as expected from a predominantly Population II background although they cannot rule out a harder one.  Future observations probing a factor of two deeper in metal column density will distinguish between the soft, fluctuating and QSOs-only UVBs.
\end{abstract}


\section{Introduction} \label{sec:intro}
The question of which sources generated the ionizing photons during the first billion year remains one of the outstanding problems in extragalactic astronomy.  For decades, it has generally been accepted that quasars were not abundant enough to provide the majority of ionizing photons before $z=3.5$~\citep{sha87} owing to the rapid evolution in the quasar luminosity function.  The alternative source, star-forming galaxies, has been forwarded repeatedly, particularly in the last decade, culminating with the recent abundance of studies leveraging measurements from \emph{HST}, \emph{Spitzer}, and other instruments~\citep{ate15,fink15, mcl13,bou15}; the general consensus is that galaxies could indeed have done the job~\citep{mad14,rob15}---under certain conditions.

Those conditions can be distilled to two major uncertainties.  The first regards the unknown abundance of faint galaxies, which could have provided the majority of star-formation sites.  Recent works have probed the rest-frame ultraviolet luminosity function (LF) of galaxies at $z=6$ to unprecedented depths, revealing a population whose abundance grows to faint luminosities with no apparent turnover~\citep{ate15,liv16}.  Reasonable extrapolations indicate that there is still room for even fainter galaxies, as also suggested by an analysis of stellar populations in nearby dwarfs~\citep{wei14}.  The brighter of these systems will be amenable to direct study by \emph{JWST}.  However, even \emph{JWST} will not probe fainter than absolute magnitudes of $M_{1500}=-15$, while direct observation through lensing clusters indicates that the luminosity function rises to at least -13~\citep{ala14,ate15,liv16} and galaxy formation models predict that it rises to -12 at $z>7$~\citep{wis14}.  Complementary ways to probe the level of activity in low-mass galaxies and test whether they could have provided the observed UVB are therefore (and will remain) urgently needed.  

The second major uncertainty in the ``galaxies-only" hypothesis is the fraction $\fesc$ of ionizing photons that escape into the intergalactic medium (IGM).  Recent work indicates that as much as 20\% of all ionizing photons may have needed to escape from star-forming galaxies in order to fill out the necessary budget for cosmological reionization~\citep{rob15}, although complementary analyses have indicated that values as low as 4--6\% may be sufficient~\citep{has15, kul13}.  If the higher number is correct, then it introduces tension with measurements of $\fesc$, which tend to favor fractions of 10\% or less both at $z\sim3$~\citep{sch06,gra15,sia15} and locally~\citep{ber06,lei13,bor14,izo16}.  A requirement that $\fesc>10\%$ would therefore cast doubt on the viability of star-forming galaxies as the dominant ionizing sources (however, see~\citealt{deb16} and~\citealt{van16} for possible counterexamples).  

The aforementioned concerns remain with us despite the recent downward revision in the measured optical depth to Thomson scattering by free electrons~\citep{pla15}.  At the same time, they leave room for speculation that a different population of sources dominated ionizing photon production.  In particular,~\citet{mad15} used new measurements of the QSO abundance at high redshift~\citep{gia15} to show that they may actually have provided enough ionizing photons to match constraints on the history of reionization.  Their argument involved balancing the inferred ionizing photon production from QSOs against requirements from reionization, but it made no reference to the character of the resulting UVB.  It re-raises the question of whether there are complementary tests that could distinguish between reionization from Population II star formation in galaxies versus QSOs.

Metal absorbers have been proposed as an alternative probe both of low-mass galaxies~\citep{bec11} and of the UVB~\citep{gir97,opp09,bol11,dod13,kea14}  In the first case, they may provide our only probe of the faintest galaxies because clouds of enriched gas are visible in absorption long before their host galaxies are visible in emission.  In the second case, metal absorbers are sensitive to the UVB at energies well above the Lyman limit.  This is useful because Population II stars give rise to a generally softer UVB than QSOs, emitting significantly less flux at energies above 4 Ryd.  Hence a measurement of the UVB slope in the range 1--5 Ryd could shed insight into the nature of the sources that generated it.

This idea was invoked by~\citet{agu04}, who found that the observed pixel optical depth statistics of $\civ$ and $\siiv$ could be reproduced with the~\citet{haa01} model at $z~\sim3$.  Applying it at higher redshifts was the goal of~\citet[][hereafter OPP09]{opp09}, who showed that ensemble statistics of aligned absorbers could distinguish between several trial UVBs.  OPP09 was limited in two respects, however.  First, it contained only a crude model for UVB spatial inhomogeneity.  This has previously been shown to be important for Damped Lyman-$\alpha$ absorbers at $z\sim3$~\citep{sch06}, which immediately raises the question of whether it matters for the slightly more diffuse systems identified via $\civ$.  OPP09 modeled UVB spatial fluctuations by assuming that, at each point, the UVB owed entirely to the nearest galaxy.  This can be seen as an extreme case: In reality, the clustered nature of structure formation means that the UVB at any position receives contributions from many local galaxies.  Moreover, this model did not treat the IGM's opacity realistically.  A more accurate treatment requires three-dimensional simulations.  Second, at the time that OPP09 was published, large samples of uniformly-selected high-redshift absorbers were not yet available for comparison.

Both of these limitations have recently been alleviated owing to two advances: (1) Publication of large data sets of metal absorbers selected both on high- and on low-ionization lines~\citep{bec11,dod13}; and (2) continuing improvements in cosmological simulations, which now include treatments for radiation transport effects in multiple independent frequency bins that source inhomogeneity in the UVB on small scales.  Leveraging these advances into an improved understanding of the reionization-epoch UVB and the sources that generated it is the goal of the present work.  In particular, we will use comparisons between a recent simulation and observations of $z\sim6$ metal absorbers to ask several questions:
\begin{itemize}
\item Do observations accommodate or require spatial fluctuations in the UVB?
\item Do observations prefer a harder over a softer UVB?
\item What further observations are needed in order to obtain an improved understanding of the sources that produced the $z\sim6$ UVB?
\end{itemize}

In Section~\ref{sec:sims}, we review our simulation and the techniques for generating artificial catalogs of metal absorbers for different UVBs.  In Section~\ref{sec:results}, we present our results.  Finally, in Sections~\ref{sec:disc}--\ref{sec:sum}, we discuss our results and summarize.

\section{Simulation and Analysis}\label{sec:sims}
\subsection{Hydrodynamics, Star Formation, Feedback, and Radiation Transport}\label{ssec:hydro}
Our cosmological radiation hydrodynamic simulation was run using a custom 
version of {\sc Gadget-3} (last described in~\citealt{spr05}) that
includes well-tested treatments for star formation and galactic winds
as well as a self-consistently computed ionizing radiation field.  
Our physical treatments and parameter choices are identical to those used in~\citet{fin15} (except that the simulation volume is roughly twice as large; see below); the reader is referred to Section 2 of that work for details.  

As our method for generating a realistic, inhomogeneous radiation field is a critical 
ingredient of the current work, we review it briefly.  The radiation field includes
contributions from galaxies and QSOs.  We discretize the galaxy radiation
field in space using a regular grid and in frequency using 16 independent
bins spaced evenly from 1--10 Rydbergs.  We obtain each voxel's 
emissivity from the instantaneous star formation rates of its gas 
particles in a way that accounts for their metallicities using the
{\sc Yggdrasil}~\citep{zac11} spectral synthesis code.  We neglect 
ionizing recombination radiation.

We adopt for $\fesc$ the same model as 
in~\citet[][Equation 1]{fin15}, in which $\fesc$ varies 
with halo mass and redshift but not frequency.  This model is tuned 
to release enough ionizing photons to yield a reasonably timely
reionization history without overproducing the post-reionization
radiation field.  $\fesc$ does not depend on frequency, with the 
result that galaxies do release photons with energies above 4 Ryd,
allowing them to ``jump-start" $\heii$ reionization.  Unfortunately,
stellar emissivities at these energies are not well-constrained, 
hence the galaxy radiation field above 4 Ryd must be
regarded as one of an ensemble of possibilities.

The emissivity 
from gas particles that straddle cell boundaries is distributed using the 
same smoothing kernel that is used to compute their hydrodynamic 
properties.  We obtain each voxel's opacity from the local abundance 
of $\hi$, $\hei$, and $\heii$.  At each timestep, we then use the moments 
of the radiation transport equation to evolve the radiation field, 
iterating to convergence between the cooling/ionization and 
radiation solvers.  
We close the moment hierarcy using Eddington tensors
that are updated frequently via a time-independent ray-casting 
calculation; see~\citet{fin11} and~\citet{fin12} for details.

We assume that the QSO radiation field is spatially-uniform throughout
our simulation volume.  This is justified because the space 
density of bright quasars at $z\sim6$ is 
$\sim10^{-9}\Mpc^{-3}$~\citep{kas15}.  A typical region
is therefore $\sim1000\, \Mpc$ away from the next quasar, a distance 
that is a factor of $\sim100$ larger than our simulation volume.

We model the QSO contribution to the UVB 
using a spatially-averaged radiation transport calculation following HM12 
(see their equation 1).  We discretize the QSO field spectrally into the 
same frequency bins as the galaxy field.  The emissivity and spectral shape 
are from Equations 37 and 38 of HM12, with the modification that the
emissivity vanishes at $z>8$.  The opacity in each frequency bin is the 
volume-average over the entire simulation.  The QSO transfer calculation 
is included in the simulation's cooling/ionization iteration, hence it 
is fully coupled into the radiation hydrodynamic framework.

Our radiation transport solver's spatial resolution is too coarse to 
resolve dense regions associated with Lyman-limit systems and 
low-ionization metal absorbers.  Without further modification, it would
therefore overestimate the UVB's ability to penetrate optically-thick
regions.  In order to model self-shielding, we use a subgrid calculation 
that attenuates the QSO and galaxy radiation fields in dense 
regions realistically under the assumption that the gas is in hydrostatic 
equilibrium~\citep{sch01}.  The ``partially-hidden" gas then makes a 
reduced contribution to the opacity field.  This two-scale approach 
leads to good agreement with direct calculations of attenuation (see 
the Appendix of~\citealt{fin15}).

Our current simulation 
models a $7.5\hmpc$ volume using $2\times320^3$ particles and discretizes
the UVB into $40^3$ independent radiation transport voxels. Hence it
subtends essentially twice the cosmological volume as our previous calculation
with exactly the same mass and spatial resolution.  We generate the 
initial conditions using an~\citet{eis99} power spectrum at $z=249$.  We
compute the initial gas ionization and temperature using {\sc RECFAST}~\citep{won08}
Our adopted cosmology is one in which $\Omega_M=0.3$, $\Omega_\Lambda=0.7$, 
$\Omega_b = 0.045$, $h=0.7$, $\sigma_8 = 0.8$, and the index of the 
primordial power spectrum  $n=0.96$.

\subsection{Extracting Simulated Catalogs of Absorbers}\label{ssec:simCat}
We extract mock catalogs of metal absorbers from our simulation using the
pipeline described in Section 5.1 of~\citet{fin15}:  First, we cast 
a sightline through the volume that is oblique to its boundaries, wrapping
around the box until it subtends a Hubble velocity width of 
$5\times10^5\kms$.  We divide the sightline into $2\kms$ pixels, onto each
of which we smooth the nearby SPH particles' density, metallicity, 
temperature, and proper velocity.  Along the way, we compute the ionization 
state of each metal species assuming ionization equilibrium in a way that 
accounts for photoionization by the 
\emph{local} radiation field\footnote{In the case of homogeneous
UVB models, the local and global UVBs are identical.}, collisionial ionization, direct and 
dielectronic recombinations, and charge transfer recombination; this 
yields the intrinsic column density in each ion.  We then use a standard
approach~\citep{the98} to compute the transmission in each transition of 
interest as a function of velocity; this accounts for thermal broadening,
bulk motions, and instrumental noise typical of high-redshift 
observations.  The transitions that we consider are $\civ \lambda 1548$,
$\siiv \lambda 1394$, and $\cii \lambda 1335$. 

Having computed simulated spectra, we use {\sc autovp}~\citep{dav97} to 
identify absorption lines.  We then merge lines within $50\kms$ of each other 
into absorption ``systems", starting with the strongest lines and working
to the weakest.  Each system's velocity is computed as the column-weighted
mean of the lines that were merged to make it.  Finally, we search for
matched $\siiv$ and $\cii$ systems that lie within $50\kms$ of $\civ$, 
working from the strongest to the weakest $\civ$ systems.

\subsection{UVBs}\label{ssec:uvbs}
\begin{figure}
\centerline{
\setlength{\epsfxsize}{0.5\textwidth}
\centerline{\epsfbox{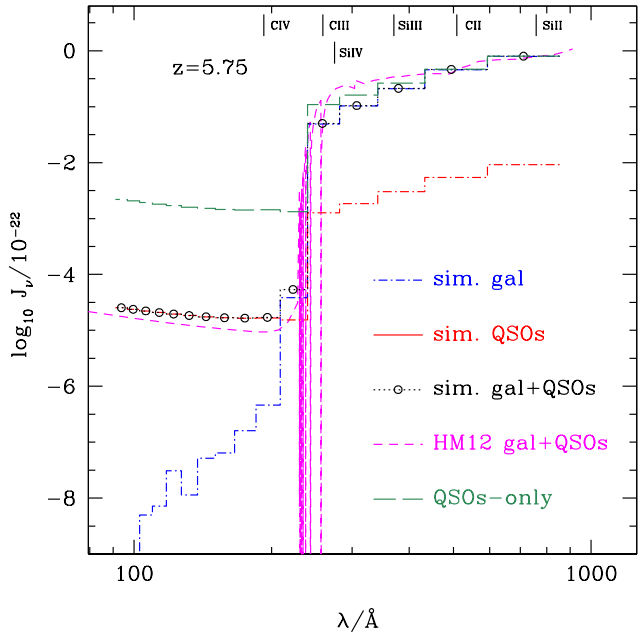}}
}
\caption{The simulated and toy-model UVBs that we consider.  The dot-dashed blue and solid red curves indicate the volume-averaged simulated galaxy and QSO UVBs, calibrated to match the~\citet{haa12} UVB in the lowest energy bin; the black dotted curve with open circles gives their sum.  The long-dashed green curve gives the QSOs-only toy-model UVB, which is the simulated QSO UVB (solid blue) calibrated so that, by itself, it matches the other UVBs in the lowest energy bin.  The magenta short-dashed curve gives the~\citet{haa12} UVB.  The photoionization potentials of relevant ions (from~\citealt{ver96}) are indicated by small ticks at the top of the figure.
}
\label{fig:uvbs}
\end{figure}

A central goal of this work is to explore how sensitive the observed population of metal absorbers is to variations in the UVB.  Hence we repeat the steps in Section~\ref{ssec:simCat} for each of three trial UVBs.  In Figure~\ref{fig:uvbs}, we compare them.

{\bf Simulated UVB} Our fiducial UVB is based on the one that is self-consistenly computed by the numerical simulation as described in Section~\ref{ssec:hydro}.  This \emph{directly simulated} UVB contains contributions from an inhomogeneous galaxy emissivity and a homogeneous QSO emissivity (Section~\ref{ssec:hydro}).  It realistically accounts for spatial fluctuations in the UVB's slope and normalization, and it regulates the predicted IGM and CGM temperatures.  As in~\citet{fin15}, however, the directly simulated UVB has a higher normalization than the~\citet[][hereafter HM12]{haa12} model, so we calibrate it so that its volume-average matches the HM12 model in its lowest energy bin (1--1.5625 Ryd).  The calibration varies with redshift but is between factors of 0.14--0.72.  This preserves the directly simulated UVB's slope and spatial inhomogeneity while rendering comparison with predictions from the HM12 UVB easier to interpret.  We shall refer to this re-calibrated version of the directly simulated UVB as the ``simulated UVB" from now on, while noting the slight departure from a completely self-consistent calculation.  The black dotted curve with open circles shows this simulated UVB.

{\bf HM12 UVB} Our next UVB is that of HM12 (magenta dashed), which we incorporate without modification.  We do not re-run our simulation with the HM12 UVB; we simply adopt it in post-processing when extracting metal absorbers.  As in~\citet{fin15}, we find that the HM12 UVB is somewhat flatter than the simulated one for energies below 4 Ryd (200--1000 \AA).  This would nominally lead the HM12 UVB to predict to more highly-ionized metals.  Opposing this, the simulated UVB incorporates local-field effects that boost the UVB's amplitude in the vicinity of galaxies.  We will see below that the latter effect dominates for metal absorbers. 

{\bf QSOs-only UVB} We obtain our final UVB by amplifying the QSO portion of the numerically simulated UVB so that it matches the other two models in the lowest energy bin, and neglecting the galaxy contribution entirely.  This is meant to represent what the simulated UVB would look like if all ionizing sources had QSO-like spectra and is shown with a green long-dashed curve.  The QSOs-only UVB has a slope that is intermediate between the simulated and HM12 UVBs for energies below 4 Ryd, but at higher energies it is significantly harder.  As in the case of the HM12 UVB, we simply adopt this trial UVB in post-processing; we do not re-run the entire simulation.  In fact, as a model for a QSOs-only UVB, this calculation probably underpredicts the flux bluewards of 4 Ryd because QSOs would ionize the IGM's $\heii$ more completely than our simulation's softer background does, suppressing the opacity at energies greater than 4 Ryd.  Accounting for this would lead to generally more highly-ionized metal absorbers.

Given that our trial UVBs do not emerge completely self-consistently from the simulation (although the simulated UVB is close), they are not consistent with the predicted gas temperatures, which we adopt from the radiation hydrodynamic simulation without modification.  In the case of the simulated and HM12 UVBs, this means that the CGM temperatures are slightly too warm, leading to slightly over-ionized gas.  The impact in the case of the QSOs-only UVB is more difficult to assess because removing local UVB amplifications would cool the CGM while steepening the UVB's slope would heat it.  In any case, we expect these effects to be slight because the ions under consideration are predicted to be predominantly photoionized, in which case their column densities should be insensitive to slight temperature changes.  Future work considering the absorbers' velocity widths may be more sensitive to the CGM temperature.

\subsection{A Sample Sightline}
\begin{figure}
\centerline{
\setlength{\epsfxsize}{0.5\textwidth}
\centerline{\epsfbox{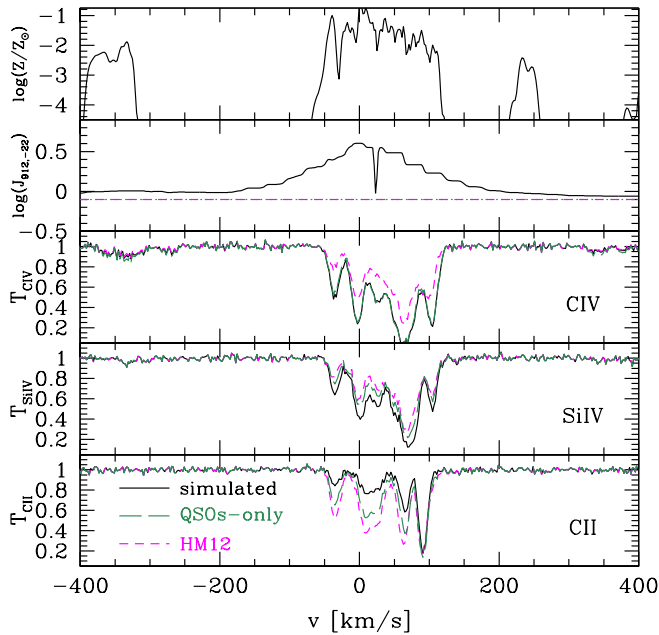}}
}
\caption{A segment of simulated sightline at $z=5.75$.  The top panel shows mass-weighted metallicity in solar units~\citep{asp09}.  The second shows the UVB amplitude in the lowest energy bin, which spans 1--1.625 Ryd (the QSOs-only and HM12 UVBs are spatially-homogeneous).  The other three panels show the calculated transmission in the $\civ1548$, $\siiv1394$, and $\cii1335$ transitions for our three UVBs. 
}
\label{fig:LOS}
\end{figure}

In order to build intuition regarding how varying the UVB changes our predictions, we explore in Figure~\ref{fig:LOS} a segment from our $z=5.75$ sightline.  In the top panel, we show mass-weighted metallicity in solar units~\citep{asp09}.  The next panel shows the strength of the simulated UVB in the lowest energy bin.  The other three panels show the predicted transmission in the three metal-ion transitions that we consider.

Comparing the top panel with the bottom three indicates that absorption is only seen where $Z/Z_\odot > 1\%$.  There are, of course, plenty of regions in which the CGM's metallicity is nonzero but lies below this level, as directly observed at lower redshifts~\citep{cri15}, but they do not produce absorption that is strong enough to be observed at $z\sim6$ with current instruments.

Comparing the UVB (second panel) with the transmission curves (bottom three panels) indicates vividly how the UVB is locally stronger throughout the region where metal absorption is found.  This reflects the presence of nearby galaxies and, as we will argue, is important for understanding the observed column density ratios.  Typically, we find much more dramatic UVB amplification near high-ionization absorbers than in areas where only low ionization metals with no high-ionization counterparts are found (see also Figure~\ref{fig:uvbFlucts}).  This is consistent with a picture in which high-ionization systems broadly favor more active neighborhoods and hence more massive host galaxies than low-ionization systems.  

In dense regions that can host $\cii$, the UVB is suppressed owing to self-shielding.  While we do account for this, we find in practice that $\cii$ is not confined to self-shielded regions.  They exist, as can be seen in the ``dip" in the UVB at $\approx30\kms$, and they do boost the $\cii$ abundance somewhat.\footnote{The dip appears very sharp because the self-shielded region has a full width at half-maximum in velocity space of $\approx4\kms$}.  On the other hand, a great deal of $\cii$ arises in a rarefied phase where the $\cii$ abundance is more effectively suppressed by the strong local UVB (because the carbon is predominantly in a more highly-ionized state) than enhanced by self-shielding.  We will see this again in Figure~\ref{fig:uvbFlucts}.

Comparing the simulated (solid black) and HM12 (dashed magenta) curves indicates that the simulated UVB broadly yields more highly-ionized metals owing to local UVB amplification.  This could owe to the simulated CGM's UVB being either more intense or harder.  We will show in Section~\ref{ssec:4Ryd} that it is in fact the local boost to the UVB's intensity that dominates.

By contrast, the QSOs-only UVB's predictions differ owing to a \emph{combination} of differences in slope and intensity.  It predicts essentially the same $\civ$ abundance as the simulated UVB because its harder slope more than compensates for the lack of local amplification.  By contrast, it predicts somewhat less $\siiv$.  This is because $\siiv$ is sensitive primarily to energies below 4 Ryd, where the two UVBs have similar slopes.  In this case, the QSOs-only model's lack of local amplifications are more important than its hardness.  For the same reason, it also predicts generally more $\cii$ absorption for this system.  

While Figure~\ref{fig:LOS} suggests that the QSOs-only UVB will predict overall more $\cii$ absorption for a given $\civ$ system, we will show in Figures~\ref{fig:civciisiiv}--\ref{fig:colDenCum} that, on average, the QSOs-only model actually \emph{under}predicts the $\cii$ column of $\civ$ absorbers.  This emphasizes that the interplay between local UVB enhancements, self-shielding, and UVB slope is nontrivial.

Figure~\ref{fig:LOS} also shows that our radiation transport solver's grid resolves UVB amplifications quite well: A close inspection of the second panel reveals a ``stairstep" shape for the local UVB.  The ``steps" reflect the radiation transport solver's grid cells, and they are clearly much smaller than the region where the UVB is enhanced.  This strongly suggests that resolution limitations associated with our radiation transport solver's coarse spatial discretization are not a serious problem.

In short, changing the adopted UVB's slope and inhomogeneity \emph{both} impact predictions regarding metal absorbers, and by considering three ions simultaneously we can disentangle their separate roles.  Differences between the HM12 and simulated UVBs will broadly reflect the role of local UVB amplitude fluctuations, while differences between the QSOs-only and simulated UVBs will reflect a combination of slope and amplitude effects.

\subsection{UVB Fluctuations}\label{ssec:uvbFlucts}
\begin{figure}
\centerline{
\setlength{\epsfxsize}{0.5\textwidth}
\centerline{\epsfbox{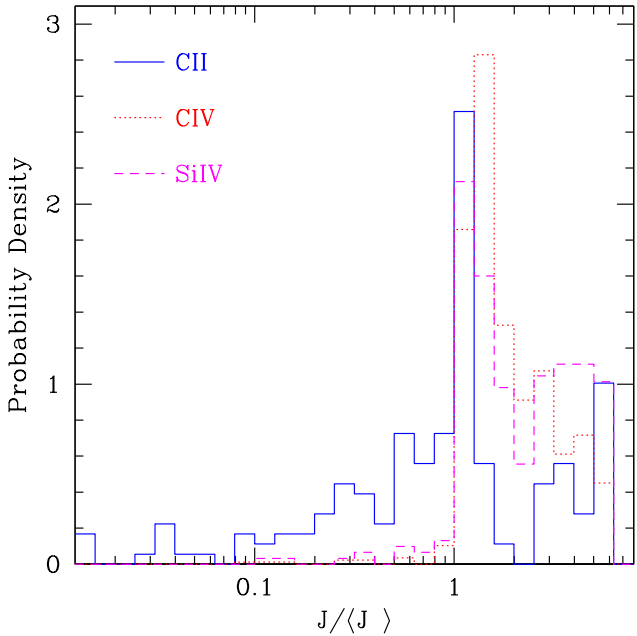}}
}
\caption{Distribution of local UVB amplitude fluctuations $J/\langle J \rangle$ in the lowest energy bin (1--1.5625 Ryd) where the column density in each of our three ions, when integrated over $2\kms$, exceeds an optical depth cutoff( see text).  A value $J/\langle J \rangle = 1$ indicates the volume-averaged mean. Results are at $z=5.75$.
}
\label{fig:uvbFlucts}
\end{figure}

A key strength of our simulation is its ability to capture small-scale UVB fluctuations that occur in the regions where metal-enriched clouds live.  We showed in~\citet{fin15} that, as expected, fluctuations on the scale of our grid cells are large at early times and smaller later on (Figure 4 of that work).  In order to quantify how significant these fluctuations are for studies of metal absorbers, we show in Figure~\ref{fig:uvbFlucts} normalized distributions of local UVB amplitude fluctuations $J/\langle J \rangle$ in positions where each ion's local column density, integrated over a velocity width of $2\kms$, exceeds an associated optical depth cutoff.  For $\civ$, this corresponds to a column density of $10^{12}$cm$^{-2}$.  For $\cii$ and $\siiv$, we scale the $\civ$ column density cutoff by the product of the oscillator strength with the central wavelength so that the $\cii$ and $\siiv$ cutoffs yield the same optical depth.  This turns out to be $10^{12.7}$ cm$^{-2}$ for $\cii$ and $10^{12.1}$ cm$^{-2}$ for $\siiv$.  The mean UVB $\langle J \rangle$ is calculated over pixels where the metallicity is zero because the distribution of local UVB amplitudes $J$ is highly non-Gaussian; however, results are essentially unchanged if we compute $\langle J \rangle$ by averaging over all pixels.  We consider only the lowest energy bin.

$\cii$ absorbers evidently trace a broad range of local UVBs, with fluctuations generally spanning factors of 0.1--5.  Those that live in obscured regions ($J/\langle J \rangle < 1$) correspond to galaxy ISM, while those that originate in locally-enhanced UVBs correspond to the CGM. By contrast, $\siiv$ and $\civ$ are found essentially exclusively in regions with $1 > J/\langle J \rangle > 6$.  A small population of high-ionization absorbers does exist in the presence of obscured UVBs and probably traces ISM gas that has been shocked by outflows.  However, inspection reveals that these are predominantly of low column densities ($\sim10^{12}$cm$^{-2}$) and are therefore not easily observable in the reionization epoch with present instrumentation.

\begin{figure}
\centerline{
\setlength{\epsfxsize}{0.5\textwidth}
\centerline{\epsfbox{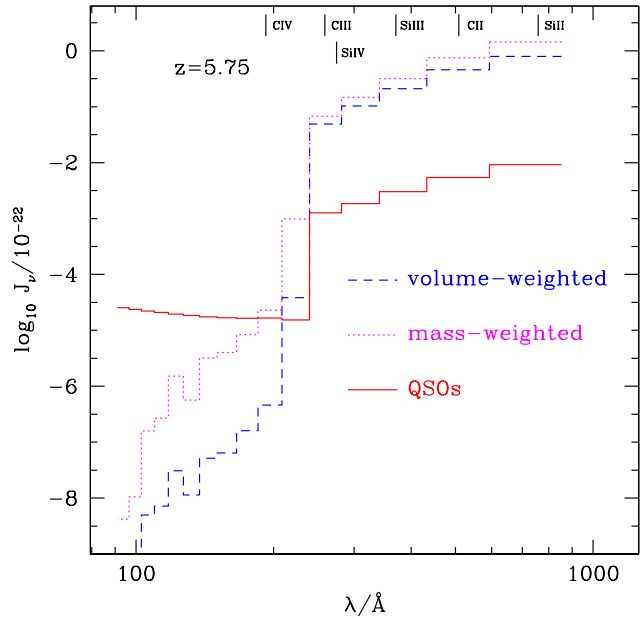}}
}
\caption{The volume- and mass-averaged galaxy UVBs in our ``simulated" model.  The QSO contribution is the same as in Figure~\ref{fig:uvbs} and is spatially-homogeneous.  Broadly, the mass--averaged UVB is harder and more intense than the volume-averaged one at $z=5.75$, indicating that the UVB which is probed by the Lyman-$\alpha$ forest is not a good indicator of the field in collapsed regions where metal absorbers live.
}
\label{fig:uvbFilt}
\end{figure}

The reionization-epoch UVB likely varied strongly in slope as well as in amplitude.  This is partly because of spectral filtering effects that would yield a harder UVB in regions farther from sources even in the presence of a uniform IGM opacity~\citep{tit07}.  However, at $z\sim6$, the IGM's opacity is far from uniform because $\heii$ is incompletely reionized~\citep{wor11,wor14}.  In particular, if galaxies contribute to the UVB above 4 Ryd, then they generate regions where $\heii$ is partially reionized, leading to locally lower opacity and a harder UVB spectral slope.  We therefore expect a rather complicated dependence of UVB slope on environment.  Our radiation transport solver treats all 16 frequency bins independently, so we may readily explore how the predicted UVB slope fluctuates with environment.

To this end, we compare in Figure~\ref{fig:uvbFilt} the mass-averaged and volume-averaged UVBs ($\JM$ and $\JV$, respectively) at $z=5.75$.  We compute $\JM$ by weighting the UVB on the resolution of our radiation transport solver by the mass in each voxel's bound groups.  This calculation ignores self-shielding, hence $\JM$ and $\JV$ roughly illustrate the differences between the environments where $\civ$ and Lyman-$\alpha$ forest absorbers live.  

It is immediately apparent that $\JM > \JV$ at all wavelengths because photons originate in galaxies.  This is characteristic of galaxy-driven reionization models: If QSOs dominated the UVB, then $\JV$ would exceed $\JM$ when averaged over regions that did not contain sources.  Put differently, galaxy-driven reionization is predominantly ``inside-out" while QSO-driven reionization is ``outside-in".  Comparing the spectral slopes, $\JM$ is much harder than $\JV$ because helium is predicted to be doubly-ionized in the vicinity of galaxies but not in the diffuse IGM.  Comparing the three UVBs with the photoionization thresholds at the top of the figure indicates that, at $z\sim6$ and in the places where metal absorbers live, galaxies dominate the UVB at all wavelengths to which the ions that we consider are sensitive.

\subsection{Rest-Frame UV Luminosity Function}
\begin{figure}
\centerline{
\setlength{\epsfxsize}{0.5\textwidth}
\centerline{\epsfbox{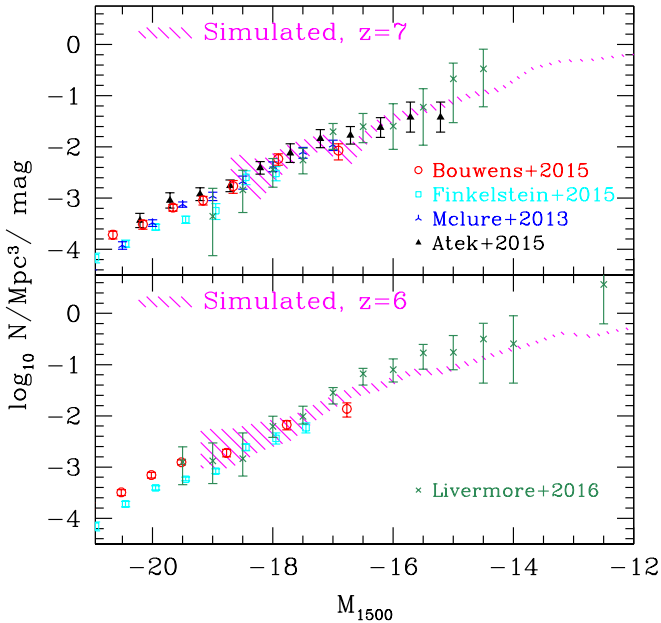}}
}
\caption{The predicted (magenta shaded region) and observed (points with errors) rest-frame UV luminosity functions at $z=7$ (top) and $z=6$ (bottom).  The shaded region indicates predicted $\sqrt{N}$ uncertainties.  We have adjusted all observations to our adopted cosmology.  The remarkably good agreement indicates that the simulation does a reasonable job of modeling the faint galaxies that host metal absorbers.
}
\label{fig:lfz6-7}
\end{figure}
One of the goals of the current work is to ask whether metal absorbers contain the signature of a fluctuating UVB that is expected of a galaxy-dominated UVB.  It is therefore important to verify that our model reproduces the abundance of the galaxies that are detectable in emission.  To this end, we compute the predicted rest-frame UV luminosity function following a standard approach.  First, we identify simulated galaxies as gravitationally-bound condensations of star and gas particles using {\sc skid}\footnote{http://www-hpcc.astro.washington.edu/tools/skid.html}.  We next compute each star particle's luminosity in an idealized 1500 \AA~filter with 15\% bandwidth and a full-width at half-maximum of 225\AA~(M.\ Dickinson, private communication) using version 2.3 of the Flexible Stellar Population Synthesis library~\citep{con09}, interpolating to the particle's metallicity and age.  Finally, we sum over each galaxy's star particles to determine the rest-frame UV luminosity.  We do not model dust extinction, which is expected to be quite minor for the luminosities and redshift ranges that we consider~\citep{bou12}.

In Figure~\ref{fig:lfz6-7}, we compare our predicted rest-frame UV luminosity functions at $z=7$ and 6 against observations by ~\citet{ate15},~\citet{fink15},~\citet{mcl13},~\citet{bou15}, and~\citet{liv16}.
While all observations indicate an encouraging level of agreement, the~\citet{ate15} and~\citet{liv16} results are particularly important tests of the model because metal absorbers are likely to trace gas associated with fainter galaxies than emission-selected samples~\citep{bec11}.  The excellent agreement at faint luminosities indicates that our simulation yields a broadly realistic galaxy population for generating and ionizing metal absorbers.

\section{Results}\label{sec:results}

\subsection{Column Density Distributions}
In this Section, we compare the predicted and observed column density distributions (CDDs) of $\civ$, $\cii$, and $\siiv$.  For each ion, we will present the predicted CDD for each of our three UVBs.  These comparisons serve as a first-order test of model and indicate to what extent the CDDs themselves probe the UVB. 

\begin{figure}
\centerline{
\setlength{\epsfxsize}{0.5\textwidth}
\centerline{\epsfbox{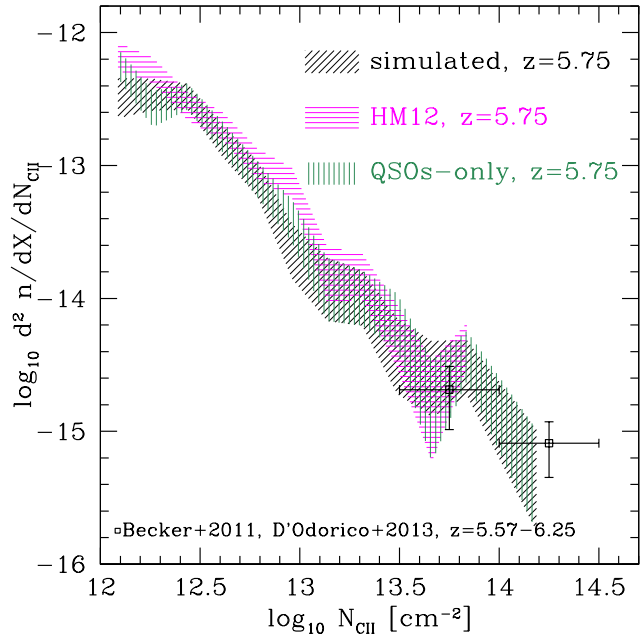}}
}
\caption{Observed $\cii$ column density distribution as compared to predictions using our three UVBs.  The widths of the simulated swaths indicate Poisson uncertainties.  All three UVB models match the data, hence the $\cii$ CDD by itself cannot discriminate between UVBs.  Observational uncertaintes are $\sqrt N$.
}
\label{fig:dndX_CII}
\end{figure}

In Figure~\ref{fig:dndX_CII}, we compare the predicted $\cii$ CDD versus observations~\citep{bec11,dod13}; see~\citet{fin15} for an explanation of how the observed CDD was compiled.  Remarkably, all three UVBs are able to match the observed $\cii$ abundance within the errors.  This is somewhat surprising given that our simulation's ability to attenuate the UVB in dense regions might have been expected to boost the $\cii$ abundance with respect to spatially-homogeneous UVBs.  In fact, this result addresses the question of whether different ions can even be used to trace ``the" UVB given that they arise in different environments (Figure~\ref{fig:uvbFlucts}).  The current set of transitions can, because much of the $\cii$ originates in a somewhat rarefied phase where self-shielding is unimportant (Section~\ref{ssec:uvbFlucts}).

\begin{figure}
\centerline{
\setlength{\epsfxsize}{0.5\textwidth}
\centerline{\epsfbox{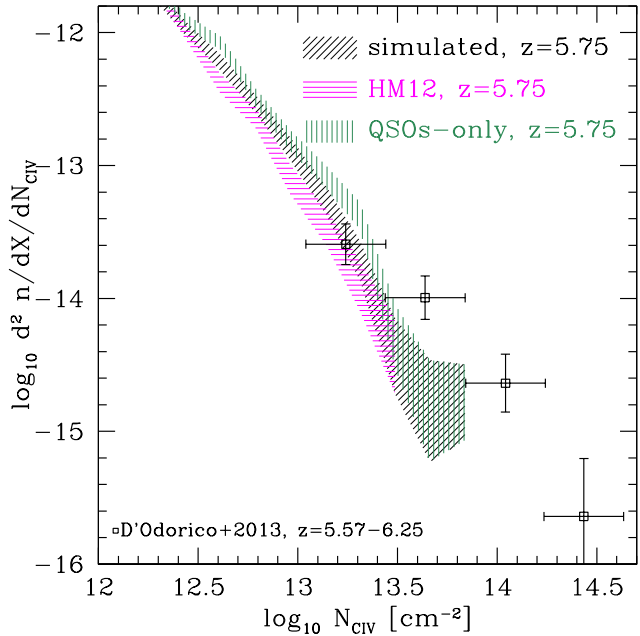}}
}
\caption{$\civ$ column density distribution at $z=5.75$ in observations~\citep{dod13} versus predictions.  All UVBs give results that are within $1\sigma$ of observations or better.  As in Figure~\ref{fig:dndX_CII}, this indicates that the $\civ$ CDD by itself cannot constrain the UVB's slope or inhomogeneity.
}
\label{fig:dndX_CIV}
\end{figure}

In Figure~\ref{fig:dndX_CIV}, we compare the observed $\civ$ CDD versus predictions.  Results are much as in~\citet{fin15}: at $10^{13.2}$cm$^{-2}$, we find excellent agreement with observations regardless of the UVB.  In detail, the QSOs-only UVB predicts the most $\civ$ owing to its harder ionizing continuum while the HM12 UVB predicts the least owing to its lack of local UVB amplifications near galaxies.  These differences are not large compared to the uncertainties, however, hence they cannot be used to distinguish the UVBs.  The systematic deficiency of strong $\civ$ absorbers in all three UVBs could indicate that even the simulated UVB underpredicts local amplifications to the UVB intensity, which if true would highlight an unexpected computational challenge.  Unfortunately, it is not easy to rule out the more mundane interpretation that our simulation lacks strong absorbers because they are associated with rare overdensities that we miss owing to our small simulation volume.  In the light of Figure~\ref{fig:lfz6-7}, this would mean that $\civ$ absorbers with column densities above $10^{13.5}$cm$^{-2}$ trace gas associated with galaxies brighter than $M_{1500}=-19$.  Indeed, the predicted CDD's slope around $10^{13}$cm$^{-2}$, if extrapolated, would likely be consistent with observations.  A more detailed study of the CDD's slope and its sensitivity to local amplifications within a simulation that treats a larger dynamic range is obviously in order.  In the meanwhile, however, we conclude that the observed $\civ$ CDD does not rule out our model and proceed to $\siiv$.

\begin{figure}
\centerline{
\setlength{\epsfxsize}{0.5\textwidth}
\centerline{\epsfbox{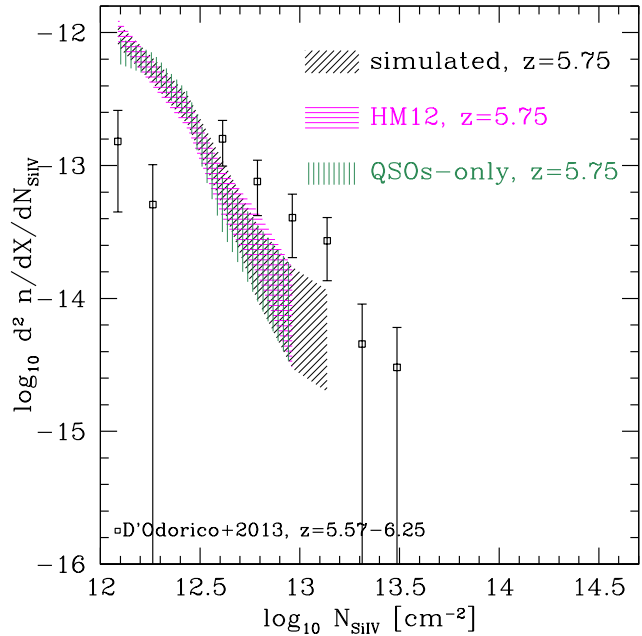}}
}
\caption{$\siiv$ column density distribution at $z=5.75$ in observations~\citep{dod13} versus predictions.  The comparison suggests observational incompleteness at columns below $10^{12.5}$cm$^{-2}$, and a systematic underproduction in the model at higher columns, qualitatively similar to Figure~\ref{fig:dndX_CIV}.  Observational uncertainties are $\sqrt N$.
}
\label{fig:dndX_SiIV}
\end{figure}

As~\citet{dod13} did not compute a $\siiv$ CDD, we compile one from their data as follows:  First, we compute the total theoretical $\civ$ and $\siiv$ path lengths for each QSO in~\citet{dod13} assuming that it is observable from a position 5000$\kms$ bluewards of the absorption line in the QSO rest-frame until 1216\AA.  Next, we assume that the accessible fraction of the total path is the same for $\civ$ and $\siiv$ and use the reported actual $\civ$ path lengths to estimate the accessible $\siiv$ path length; this comes out to $dX=21.6608$ for our cosmology.  Next, we compile the 25 $\siiv$ absorbers for which column densities are reported and ignore those for which only upper limits or equivalent widths are reported.  Finally, we estimate uncertainties using $\sqrt{N}$.  Neglecting the three blended/contaminated systems for which~\citet{dod13} report only the equivalent widths may lead us to underestimate the true CDD's amplitude by $\approx10\%$.

The result, which we show in Figure~\ref{fig:dndX_SiIV}, is that our model overproduces the observed CDD below $10^{12.5}$cm$^{-2}$ and underproduces it at $10^{12.5}$--$10^{13.25}$cm$^{-2}$.  We attribute the overproduction at low columns to observational incompleteness.  At higher columns, the level of disagreement and the tendency for the model to underproduce observations just above a point where the predicted CDD turns over bears a qualitative resemblance to the $\civ$ CDD's behavior (Figure~\ref{fig:dndX_CIV}), and we attribute it once again to the lack of rarer, somewhat more massive galaxies that likely host $\siiv$ absorbers.

In summary, Figures~\ref{fig:dndX_CII}--\ref{fig:dndX_CIV} indicate that low-ionization absorbers are well-reproduced while high-ionization absorbers are underproduced, particularly at high columns.  These comparisons support the view (from OPP09) that low-ionization absorbers occur in relatively dense gas that lies near galaxies, and are hence dominated by low-mass host galaxies that do not expel their outflows very far.  Meanwhile, high-ionization systems are associated with rarer, more massive galaxies whose outflows are sufficiently energetic to expel gas out to the low densities where high ionization states are energetically favored.  Such systems are systematically underrepresented in our simulation even though it subtends twice the volume of the calculation considered in~\citet{fin15}.  

For the rest of this paper, we will treat all metal absorbers as an ensemble.  This approach defers to future work an inquiry into the risks inherent in comparing a model that is dominated by low-mass galaxies to data that are dominated by more massive galaxies.

\subsection{Abundance Ratios}\label{ssec:ratios}

Figures~\ref{fig:dndX_CII}, \ref{fig:dndX_CIV}, and~\ref{fig:dndX_SiIV} suggest that, subject to our numerical limitations, the simulation produces roughly the correct overall number $\cii$, $\civ$, and $\siiv$ absorbers.  This gives us confidence that the spatial distribution of metals, the thermal state of the simulated CGM,  and the UVB are broadly realistic.  We now progress from CDDs of individual ions to the statistics of aligned metal absorbers.  This serves two complementary purposes.  First, it is entirely possible that the level of agreement suggested by the CDDs masks problems in the CGM phase structure.  For example, the model could produce the correct overall amount of $\cii$ but distribute it in all the wrong places.  Such problems might reveal themselves as a discrepancy between the observed and modeled $\cii$ columns that are spatially associated with $\civ$ absorbers.  Secondly, aligned absorbers can be wielded as a probe of the UVB's inhomogeneity and hardness.  Following OPP09, we will do this by leaning on the assumption that the simulated CGM's metallicity and thermal properties are realistic, and applying different UVB models in post-processing in order to ask which one best reproduces the observations.

\begin{figure}
\centerline{
\setlength{\epsfxsize}{0.5\textwidth}
\centerline{\epsfbox{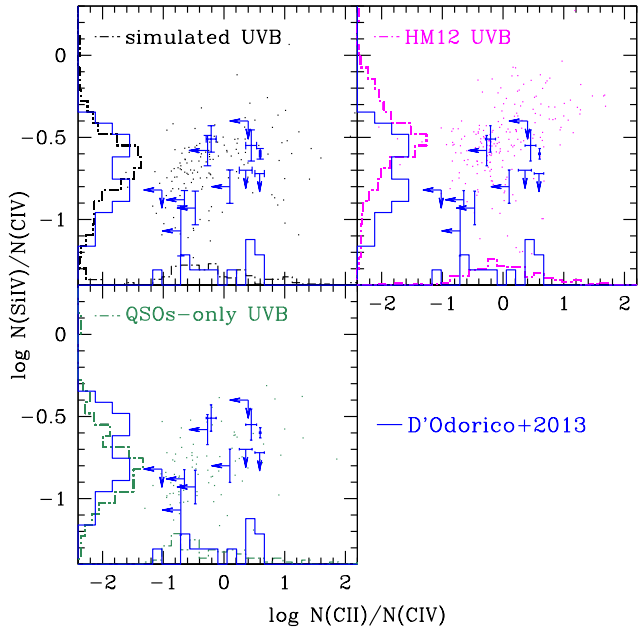}}
}
\caption{The observed abundance ratios of $\siiv/\civ$ versus $\cii/\civ$ (blue points with errors) as compared to results from our three different UVB models.  The simulated UVB (top left) yields good agreement with observations.  By contrast, the HM12 (top right) model underpredicts and the QSOs-only model (bottom left) overpredicts the observed $\civ$ abundance relative to $\cii$ and $\siiv$. 
}
\label{fig:civciisiiv}
\end{figure}

In Figure~\ref{fig:civciisiiv}, we compare the aggregated absorber populations from the $z=5.875$ and $z=5.5$ simulation snapshots with each of our three UVBs versus the observations, which are centered at $z=5.75$.  For each snapshot, all three UVBs are normalized to have the same volume-averaged amplitude in the lowest frequency bin, hence this figure illustrates the data's sensitivity to changes in slope and, in the case of the simulated UVB, amplitude.  The solid blue histograms indicate the observed distribution with limits treated as data points; they are the same in each panel.

From the top-right panel, we see that the HM12 UVB dramatically overproduces $\siiv$ with respect to $\civ$ (this is particularly clear from the histograms), while the $\cii/\civ$ ratios show plausible agreement.  Either the HM12 UVB is too soft or weak, or else it lacks local fluctuations associated with the galaxies that create it.  Given the similarity between the simulated and HM12 UVBs (Figure~\ref{fig:uvbs}), we favor the conclusion that spatial inhomogeneity is the missing ingredient (we will confirm this in Section~\ref{ssec:4Ryd}).

In the top-left panel, we show that the simulated UVB yields excellent agreement with the observations.  It may be that the observations indicate a somewhat broader spread in the distribution of $\siiv/\civ$ ratios, but an origin in observational uncertainties, which are incompletely accounted for in the model, cannot be ruled out.

In the bottom-left panel, we see that the QSOs-only UVB predicts an absorber population with \emph{too much} $\civ$ on both axes; it is in this sense the opposite of the HM12 UVB.  Modifying the QSOs-only UVB to account for local amplifications (such as might be expected if the UVB originated in stars that sample a top-heavy initial mass function) would exacerbate the disagreement.  Hence, if we accept that the underlying model for the CGM is realistic, then the dominant ionizing sources at $z>5$ must yield a UVB whose slope is softer than the QSOs-only model in Figure~\ref{fig:uvbs}.

\begin{figure}
\centerline{
\setlength{\epsfxsize}{0.5\textwidth}
\centerline{\epsfbox{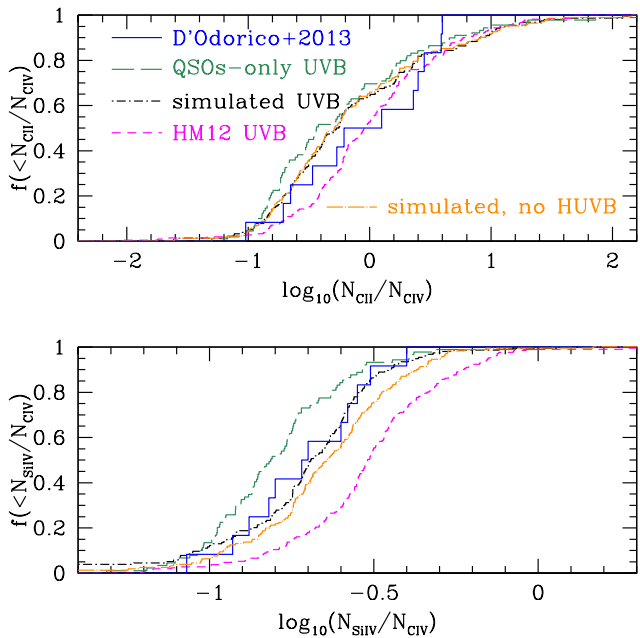}}
}
\caption{Cumulative abundance of $\civ$ absorber versus column density ratio for our three trial UVBs.  Additionally, we consider a model in which photons with energies $> 4$ Ryd are neglected (orange dot-long-dashed).  Broadly, the simulated UVB (black dot-short-dashed) performs the best in comparison to observations (solid blue); this is particularly clear in the bottom panel.
}
\label{fig:colDenCum}
\end{figure}

While the histograms in Figure~\ref{fig:civciisiiv} suggest that the simulated UVB outperforms the alternatives, it is difficult to get from them an overall sense for the qualities of the matches, and how they compare.  For clarity, we therefore redisplay the observed and predicted $\siiv/\civ$ and $\cii/\civ$ abundance ratios using cumulative distributions in Figure~\ref{fig:colDenCum}.  The QSOs-only UVB clearly overproduces $\civ$ with respect to $\cii$ and $\siiv$, indicating that it is too hard.  The HM12 model dramatically underproduces $\siiv$, but its performance in the $\cii/\civ$ space is more subtle.  It underproduces the fraction of highly-ionized $\civ$ systems ($N_\cii/N_\civ < 1$) while faring comparably to the other models in reproducing the abundance of neutral systems ($N_\cii/N_\civ > 3$).  The discrepancy may reflect the lack of local amplifications in the HM12 model, which causes it to underestimate the ionization of rarefied gas that lives near galaxies.

In contrast to the homogeneous UVBs, the simulated UVB reproduces the distribution of $\siiv/\civ$ abundance ratios stunningly well.  It shows a slight tendency to overproduce the abundance of moderately neutral $\civ$ absorbers ($N_\civ \sim N_\cii$), but is still a better match than the QSOs-only model.  If confirmed with future data, then the $\cii/\civ$ abundances could suggest that the simulation's self-shielding treatment is too efficient.

Interestingly, all three UVB models overproduce the observed abundance of highly-neutral $\civ$ systems ($N_\cii / N_\civ > 3$).  This can easily be attributed to the fact that the simulation does not convert very dense carbon into CI or CO, as happens in reality.  If this is confirmed with future data, it has the surprising implication that the statistics of $\cii$ absorbers are sensitive to the physics of molecular cloud formation at high redshift.

There are three caveats to bear in mind when interpreting Figure~\ref{fig:colDenCum}.  First, one could imagine an Eddington-like bias in which systems near the $\civ$ detection limit are contaminated by weak $\civ$ absorbers that are scattered into the sample owing to errors in the spectra around $\civ$.  This would generate a sub-population whose $\cii/\civ$ and $\siiv/\civ$ ratios implied an overly-strong UVB.  This is not corrected for.  Relatedly, observational uncertainties are not taken into account completely.  Finally, systems for which one of the three ions is undetected and reported as an upper limit, are treated as measurements in this analysis.  We do this because only 3 out of the 12 3-absorber systems in~\citet{dod13} have measurements for all three ions.  If, however, future observations indicate \emph{significantly} less $\cii$ or $\siiv$ in those cases where upper limits are currently available, then the comparison could favor the QSOs-only model.

\begin{table}
\caption{Kolmogorov-Smirnov $p$-values.}
\centering
\begin{tabular}{l | c c c}
\hline
ratio & simulated & QSOs-only & HM12 \\
\hline
\hline
measured upper limits & & & \\
$\cii/\civ$ & 0.79 & 0.47 & 0.82 \\
$\siiv/\civ$ & 0.43 & 0.20 & 0.022 \\
\hline
upper limits divided by 2 & & & \\
$\cii/\civ$ & 0.42 & 0.78 & 0.31 \\
$\siiv/\civ$ & 0.056 & 0.75 & 0.0032\\
\hline
\end{tabular}
\label{tab:pvalues}
\end{table}

We may explore the potential significance of future, deeper data by dividing all of the column density upper limits by two and comparing the resulting Kolmogorov-Smirnov $p$-values; these are reported in Table~\ref{tab:pvalues}.  The $p$-value is the probability that the simulated and observed cumulative histograms would be more different than they are if they were drawn from the same underlying distribution.  $p$-values larger than 10\% are conventionally regarded as indicating a plausible level of consistency.  Looking at the top table first confirms the visual impressions from Figure~\ref{fig:colDenCum}: The HM12 model is ruled out by the observed $\siiv/\civ$ ratios while the simulated UVB is preferred to the QSOs-only UVB for both ions.  If, however, future measurements revise all of the reported $\siiv$ and $\cii$ upper limits down by a factor of two, then the bottom table shows that the QSOs-only model would be the only one with an acceptable $p$-value (that is, $>10\%$) for both ratios.

\subsection{Harder or More Intense?}\label{ssec:4Ryd}
Figure~\ref{fig:colDenCum} raises the question of whether the UVB experienced by metal absorbers is harder than the HM12 UVB, more intense, or both.  In order to address this question, we will focus on the role of photons with energies in excess of 4 Ryd.  For simplicity, we will refer to these as HUV photons (for hard UV), and the background abundance of HUV photons as the HUVB.  

The HUVB at $z\sim6$ is particularly challenging to model for two reasons.  First, the contribution from stars is largely unknown \emph{a priori}, leading to uncertainty in our adopted galaxy emissivities.  Second, the contribution from quasars probably varied dramatically in space because the reionization of $\heii$ (which dominates the HUV opacity) was incomplete~\citep{wor11,wor14}.  If the HUVB was generated entirely by quasars, then optically-thin patches were increasingly rare and widely-separated prior to $z=3$.  Alternatively, if galaxies were bright in HUV, then the HUVB was high near both galaxies and quasars, and $\heii$ reionization was a more protracted process.  Reionization-epoch metal absorbers are a uniquely-suitable probe of the HUV emissivity from young stars and of the role that galaxies may have played in $\heii$ reionization: Given that the typical metal absorber is not conventionally expected to live in a quasar-produced $\heiii$ region (\citealt{mcq09}, but see~\citealt{mad15}), abundance ratios that indicate an HUVB would essentially require a stellar contribution.

In order to explore this idea, we have performed an additional test in which we re-extracted our sightlines using our re-normalized, simulated UVB but with the HUVB neglected (both from QSOs and from galaxies).  This no-HUV background mimics what would be expected in a region where $\heii$ was not yet reionized and assuming that stars do not produce HUV photons. We compare the result with observations and with our simulated UVB using the dark orange dot-long-dashed curve in Figure~\ref{fig:colDenCum}.  The no-HUV UVB produces essentially the same $\cii/\civ$ abundance ratios, indicating that they are relatively insensitive to the HUVB; we have also verified that the $\cii$ and $\civ$ CDDs are unaffected.

By contrast, omitting the HUVB boosts the overall $\siiv$ abundance slightly, improving agreement with the observed $\siiv$ CDD (not shown) while degrading the agreement with the $\siiv/\civ$ abundance ratios.  With significantly improved measurements and models it may therefore be possible to use $\siiv$ to study the galaxy contribution to the HUVB, but at the present its effect is too weak for precise constraints.  Given the similarity between the HM12 and simulated UVBs redwards of 4 Ryd (Figure~\ref{fig:uvbs}), and the fact that fluctuations in the simulated UVB's slope present mostly as a variation in the $\heii$ opacity (Figure~\ref{fig:uvbFilt}), this means that $\civ,\cii,{\rm\, and\,} \siiv$ are primarily sensitive to the UVB's amplitude fluctuations.

In summary, the $\cii/\civ$ and $\siiv/\civ$ abundance ratios of~\citet{dod13} indicate a UVB that is more intense than in the HM12 model.  The data are consistent with the CGM's UVB having been somewhat harder as well, although this does not seem to be a dominant effect.  The UVB was in any case not as hard (on average) as in a QSOs-only model (or, equivalently, as it would be if the metal absorbers all happened to originate near a quasar).  Current observations therefore disfavor reionization models in which the majority of the UVB at $z\sim6$ owes to QSOs.

\section{Discussion}\label{sec:disc}
This work has explored whether measurements of metal absorbers constrain two different properties of the UVB, namely its spatial fluctuations in amplitude and hardness.  In a sense, we ask whether observations require a UVB that has spatial fluctuations (unlike the HM12 model), and whether they permit an overall harder UVB, as would be expected if QSOs produced most of the ionizing photons.  We frame our discussion by addressing these questions in turn.

The significance of localized UVB enhancements near galaxies can be appreciated by comparing how well the HM12 and simulated UVB's reproduce the $\siiv/\civ$ abundance ratios.  Both ions trace optically-thin gas, hence they are not influenced by the uncertain nature of self-shielding on small scales.  Given that both models have roughly the same slope when averaged over large scales, their differences must trace the impact of galaxy-scale fluctuations.  The tendency for the UVB to be locally stronger and harder near galaxies in the simulated UVB naturally suppresses the $\siiv/\civ$ column density ratios into improved agreement with observations.

Our argument that metal absorbers probe a spatially-inhomogeneous UVB rests on the assumption that the HM12 model is accurate in the diffuse IGM; that is, far from galaxies and quasars.  Support for this comes from observations of the Lyman-$\alpha$ forest, which imply a fairly uniform UVB at the Lyman limit by $z\approx 5.6$~\citep{bec15}.  Additional support comes from $\oi$ absorbers, whose abundance at $z\sim6$ is consistent with the same photoionization rate that is inferred from the Lyman-$\alpha$ forest~\citep{kea14}.  However, if future observations indicate an overall harder and more intense volume-averaged UVB, then the need for spatial fluctuations will be correspondingly weaker.

One might suppose that the HM12 model could be forced to match the observed $\siiv/\civ$ abundance ratios by changing the assumed metal yields (see also~\citealt{dod13}).  For example, assuming that reionization-epoch SNe produce $\approx50\%$ less silicon than in our model (without changing the carbon yield) would suppress the HM12 model's $\siiv/\civ$ ratios into agreement with observations.   Unfortunately, this adjustment would exacerbate the gap between the predicted and observed $\siiv$ CDDs in Figure~\ref{fig:dndX_SiIV}.  Moreover, it conflicts with observational indications that the high-redshift ratio of carbon to silicon is, if anything, suppressed with respect to solar~\citep{bec12}.  For these reasons, we consider an origin in an incorrect silicon yield to be ruled out.

The need for self-shielding is not as well-demonstrated by these comparisons because much of the $\cii$ arises in gas that is not dense enough to be significantly shielded.  If it were, then the predicted $\cii$ column density distributions for the HM12 and simulated UVBs would not be so similar (Figure~\ref{fig:dndX_CII}).  Future work considering tracers of even denser gas such as $\oi$ will be required to connect the self-shielded gas phase with the metagalactic UVB (see also~\citealt{kea14,kea16}).

Turning to the question of the UVB's overall spectral hardness, Figure~\ref{fig:colDenCum} clearly shows that a QSOs-only UVB over-ionizes the gas where metal absorbers live.  As before, one might be tempted to assume that reionization-epoch SNe produce $\approx50\%$ more silicon than in our model (without changing the carbon yield), which would boost the QSOs-only model's predictions into agreement with observations.   Conveniently, this would likewise alleviate the gap between the predicted and observed $\siiv$ CDDs in Figure~\ref{fig:dndX_SiIV}, which we otherwise attribute to numerical limitations.  However, this trick would not work for the $\cii/\civ$ distribution, which disfavors the QSOs-only model with respect to the simulated one (although weakly, because both yield $p$-values above 10\%).  Given that adding local fluctuations to the QSOs-only UVB would exacerbate the disagreement with observed $\cii/\civ$ ratios, rescue lies in supposing that the CGM is somehow colder, denser, or both than in our model in order to boost $\cii/\civ$.  Such a change would inevitably boost the volume-averaged optical depth to Lyman-$\alpha$ absorption ($\tau_{\mathrm{Ly}\alpha}$) as well.  We showed in~\citet{fin15} that, with the renormalized UVB, our model already yields a $\tau_{\mathrm{Ly}\alpha}$ that lies nearer the high end of what is allowed observationally (Figure 2 of that paper), so this is not favored either.  Our preferred interpretation is therefore that the QSOs-only model fails to reproduce the~\citet{dod13} measurements because it is too hard.

While we have not specifically addressed predictions in the case of a model where the majority of ionizing photons owe to massive, metal-free ``Population III" stars, we may comment briefly on what might be expected.  The slope of the ionizing continuum from Population III stars is expected to be harder than that of Population II stars, and the UVB's spatial fluctuations would be comparable to what occurs in our simulation.  Hence a Population III-only model would yield generally more ionized $\cii/\civ$ and $\siiv/\civ$ ratios than our simulated UVB, increasing the tension with observations in Figure~\ref{fig:colDenCum}.  If the intrinsic slope of the Population III ionizing continuum is comparable to that of QSOs, and if we assume that the CGM's metallicity would otherwise be unchanged, then such a model would be in even \emph{poorer} agreement with observations than our QSOs-only model owing to local UVB amplifications near galaxies.  Note that this is equivalent to a scenario in which reionization was driven by faint, low-mass QSOs living in galaxies that are at or below current detection limits; such a model is therefore likewise in doubt.  In detail, supernovae from very massive stars produce more silicon per carbon than Population II IMFs~\citep[][Table 3]{kul13}, which would somewhat alleviate the tension with the observed $\siiv/\civ$ ratios.  However, absent a more detailed calculation and given that the $\cii/\civ$ ratios would almost certainly be in increased tension with the data, we conclude that the~\citet{dod13} measurements exert pressure on Population III-driven reionization scenarios~\citep[see also][]{bec11,kul13}.

\section{Summary}\label{sec:sum}
We have combined predictions from a radiation-hydrodynamic simulation with observations of three-ion metal aborber systems to study the slope and inhomogeneity of the UVB at the close of the reionization epoch ($z\sim6$).  We post-process snapshots from our simulation with three radiation fields: The ``simulated" UVB, which has a soft spectral slope but is both harder and more intense near galaxies; a hard-and-homogeneous ``QSOs-only" UVB in which it is assumed that hard sources dominate the UVB; and the soft-and-homogeneous HM12 UVB.  At a given redshift, all three UVBs are calibrated to have the same volume-averaged amplitude in our lowest frequency bin (1--1.5625 Ryd).  Our results are as follows:
\begin{itemize}
\item Our model predicts that $\civ$ and $\siiv$ originate in regions where the UVB is locally 1--6 times stronger than the volume average, while a fair fraction of all $\cii$ originates in regions that are dense enough to be self-shielded, leading to a suppressed local UVB.
\item All three UVBs fare comparably in their ability to reproduced the observed $\cii$, $\civ$, and $\siiv$ column density distributions.  $\cii$ is particularly well-reproduced, indicating that it traces low-mass galaxies whose abundance is well-represented.  $\siiv$ and $\civ$ are underproduced at high columns, consistent with the view that the simulation volume systematically underproduces the large galaxies that can expel gas to low densities where high ionization states are energetically favored.
\item The simulated UVB matches the observed distribution of $\siiv/\civ$ ratios startlingly well and fares somewhat better than the other models in reproducing $\cii/\civ$ (as quantified by their $p$-values).
\item The QSOs-only UVB overproduces $\civ$ with respect to both $\siiv$ and $\cii$, suggesting that it is too hard.  Current observations therefore disfavor reionization models in which QSOs dominate the UVB at $z\sim6$ (for example,~\citealt{mad15}).
\item The HM12 UVB clearly overproduces $\siiv$ at the position of $\civ$ absorbers, but it reproduces the $\cii$ column densities nearly as well as the simulated UVB.  Given that the HM12 and simulated UVBs have similar slopes redwards of 4 Ryd, and that omitting photons with energies $> 4$ Ryd (the ``HUVB") has little impact on results, this implies that metal absorbers probe small-scale amplitude fluctuations in the UVB.
\item $\siiv/\civ$ is a weak probe of the HUVB and in fact prefers models in which an HUVB is present near low-mass galaxies at $z\sim6$.  However, both observations and models will need to be improved to draw robust conclusions.
\end{itemize}

There are three sources of uncertainty that will need to be revisited in future work.  First, this work makes the assumption that the simulated distribution of densities, metallicities, and temperatures with the CGM of $z\sim6$ galaxies is realistic.  The plausible level of agreement between the simulated and observed CDDs supports this view, but future work considering complementary constraints such as a larger number of ions and the absorbers' velocity widths will be needed to test it further.  

Second, the relatively limited simulation volume, a trade-off for high mass resolution, obviously leaves in question whether the simulated absorbers' host systems are a good analogue for the observed absorbers'.  In the case of systems selected as low-ionization absorbers, the $\cii$ CDD indicates that this is reasonable.  However,~\citet{dod13} select on $\civ$, which is biased toward more massive host systems, in which our current model is systematically deficient.  Accounting more completely for massive systems would boost the abundance of high-column $\siiv$ and $\civ$ absorbers, irrespective of the UVB.  In the case of the simulated UVB, it would additionally lead to an overall more highly-ionized absorber population because massive galaxies live in regions with a more intense UVB.

Relatedly, our work suffers from the fact that the QSO UVB is not spatially-homogeneous prior to reionization.  A significantly larger dynamic range will be required to treat the QSO field and its impact on metal absorbers realistically.

By far our biggest source of uncertainty, however, is the fact that only three of the 12 observed systems have measurements in all three ions; we are forced to treat limits in the other systems as measurements in order to yield a statistically meaningful sample.  There is clearly an argument for obtaining deeper measurements of existing sightlines, because pushing to lower columns will be necessary in order to assess whether the systems that have already been identified trace a significantly harder UVB than current limits indicate.  However, expanding the sample to include more sightlines would likewise be helpful because sightlines that support good measurements in three different ions are rare.

\section*{Acknowledgements}
We thank J.\ Hennawi, M.\ Prescott, R.\ Simcoe, G.\ Becker, E.\ Ryan-Webber, 
R.\ Cen, and C.\ Churchill for helpful conversations and input.  We are 
indebted to V.\ Springel for making {\sc Gadget-3} available to our group, 
and to P.\ Hopkins for kindly sharing his SPH module.  We thank for the anonymous referee 
for pointing out a number of opportunities for clarification that improved the draft.  
Our simulation was run on {\sc Gardar}, a joint Nordic supercomputing facility in Iceland.
KF thanks the Danish National Research Foundation for funding the Dark Cosmology Centre.  
EZ acknowledges research funding from the Swedish Research Council, the Wenner-Gren 
Foundations and the Swedish National Space Board.


\begin{thebibliography}{99}
\bibitem[Aguirre et al.(2004)]{agu04} Aguirre, A., Schaye, 
J., Kim, T.-S., et al.\ 2004, \apj, 602, 38 
\bibitem[Alavi et al.(2014)]{ala14} Alavi, A., Siana, B., 
Richard, J., et al.\ 2014, \apj, 780, 143
\bibitem[Asplund et 
al.(2009)]{asp09} Asplund, M., Grevesse, N., Sauval, A.~J., \& Scott, P.\ 2009, \araa, 47, 481 
\bibitem[Atek et al.(2015)]{ate15} Atek, H., Richard, J., 
Jauzac, M., et al.\ 2015, arXiv:1509.06764
\bibitem[Becker et al.(2011)]{bec11} Becker, G.~D., Sargent, 
W.~L.~W., Rauch, M., \& Calverley, A.~P.\ 2011, \apj, 735, 93 
\bibitem[Becker et al.(2012)]{bec12} Becker, G.~D., Sargent, 
W.~L.~W., Rauch, M., \& Carswell, R.~F.\ 2012, \apj, 744, 91 
\bibitem[Becker et al.(2015)]{bec15} Becker, G.~D., Bolton, 
J.~S., Madau, P., et al.\ 2015, \mnras, 447, 3402
\bibitem[Bergvall et 
al.(2006)]{ber06} Bergvall, N., Zackrisson, E., Andersson, B.-G., et al.\ 2006, \aap, 448, 513
\bibitem[Bolton et al.(2011)]{bol11} Bolton, J.~S., Haehnelt, 
M.~G., Warren, S.~J., et al.\ 2011, \mnras, 416, L70
\bibitem[Borthakur et al.(2014)]{bor14} Borthakur, S., 
Heckman, T.~M., Leitherer, C., \& Overzier, R.~A.\ 2014, Science, 346, 216
\bibitem[Bouwens et al.(2012)]{bou12} Bouwens, R.~J., 
Illingworth, G.~D., Oesch, P.~A., et al.\ 2012, \apj, 754, 83
\bibitem[Bouwens et al.(2015)]{bou15} Bouwens, R.~J., 
Illingworth, G.~D., Oesch, P.~A., et al.\ 2015, \apj, 803, 34 
\bibitem[Conroy et al.(2009)]{con09} Conroy, C., Gunn, J.~E., 
\& White, M.\ 2009, \apj, 699, 486
\bibitem[Crighton et al.(2015)]{cri15} Crighton, N.~H.~M., 
O'Meara, J.~M., \& Murphy, M.~T.\ 2015, arXiv:1512.00477
\bibitem[D'Odorico et al.(2013)]{dod13} D'Odorico, V., 
Cupani, G., Cristiani, S., et al.\ 2013, \mnras, 435, 1198
\bibitem[Dav{\'e} et al.(1997)]{dav97} Dav{\'e}, R., 
Hernquist, L., Weinberg, D.~H., \& Katz, N.\ 1997, \apj, 477, 21
\bibitem[de Barros et al.(2016)]{deb16} de Barros, S., Vanzella, E., Amor{\'{\i}}n, R., et al.\ 2016, \aap, 585, A51 
\bibitem[Eisenstein \& Hu(1999)]{eis99} Eisenstein, D.~J., \& Hu, W.\ 1999, \apj, 511, 5 
\bibitem[Finkelstein et al.(2015)]{fink15} Finkelstein, S.~L., 
Ryan, R.~E., Jr., Papovich, C., et al.\ 2015, \apj, 810, 71
\bibitem[Finlator et al.(2011)]{fin11} Finlator, K., 
Dav{\'e}, R., \"{O}zel, F.\ 2011, \apj, 743, 169
\bibitem[Finlator et al.(2012)]{fin12} Finlator, K., Oh, 
S.~P., {\"O}zel, F., \& Dav{\'e}, R.\ 2012, \mnras, 427, 2464
\bibitem[Finlator et al.(2015)]{fin15} Finlator, K., 
Thompson, R., Huang, S., et al.\ 2015, \mnras, 447, 2526
\bibitem[Giallongo et 
al.(2015)]{gia15} Giallongo, E., Grazian, A., Fiore, F., et al.\ 2015, \aap, 578, A83
\bibitem[Giroux \& Shull(1997)]{gir97} Giroux, M.~L., \& Shull, J.~M.\ 1997, \aj, 113, 1505
\bibitem[Grazian et al.(2015)]{gra15} Grazian, A., Giallongo, 
E., Gerbasi, R., et al.\ 2015, arXiv:1509.01101
\bibitem[Haardt 
\& Madau(2001)]{haa01} Haardt, F., \& Madau, P.\ 2001, XXXIst Moriond Astrophysics Meeting, Clusters of Galaxies
and the High Redshift Universe Observed in X-Rays, ed. D. M. Neumann \& J. Tr{\^a}nh Thanh V{\^a}n (astro-ph/0106018)
\bibitem[Haardt 
\& Madau(2012)]{haa12} Haardt, F., \& Madau, P.\ 2012, \apj, 746, 125 
\bibitem[Hassan et al.(2015)]{has15} Hassan, S., Dav{\'e}, 
R., Finlator, K., \& Santos, M.~G.\ 2015, arXiv:1510.04280
\bibitem[Izotov et al.(2016)]{izo16} Izotov, Y.~I., 
Orlitov{\'a}, I., Schaerer, D., et al.\ 2016, \nat, 529, 178
\bibitem[Kashikawa et al.(2015)]{kas15} Kashikawa, N., 
Ishizaki, Y., Willott, C.~J., et al.\ 2015, \apj, 798, 28
\bibitem[Keating et al.(2014)]{kea14} Keating, L.~C., 
Haehnelt, M.~G., Becker, G.~D., \& Bolton, J.~S.\ 2014, \mnras, 438, 1820
\bibitem[Keating et al.(2016)]{kea16} Keating, L.~C., Puchwein, E., Haehnelt, M.~G., Bird, S., 
\& Bolton, J.~S.\ 2016, arXiv:1603.03332
\bibitem[Kulkarni et al.(2013)]{kul13} Kulkarni, G., 
Rollinde, E., Hennawi, J.~F., \& Vangioni, E.\ 2013, \apj, 772, 93 
\bibitem[Leitet et 
al.(2013)]{lei13} Leitet, E., Bergvall, N., Hayes, M., Linn{\'e}, S., \& Zackrisson, E.\ 2013, \aap, 553, A106
\bibitem[Livermore et al.(2016)]{liv16} Livermore, R.~C., Finkelstein, S.~F.\ and Lotz, J.~M.\ 2016, \emph{in prep}
\bibitem[Madau \& Haardt(2015)]{mad15} Madau, P., \& Haardt, F.\ 2015, arXiv:1507.07678 
\bibitem[Madau 
\& Dickinson(2014)]{mad14} Madau, P., \& Dickinson, M.\ 2014, \araa, 52, 415 
\bibitem[McLure et al.(2013)]{mcl13} McLure, R.~J., Dunlop, 
J.~S., Bowler, R.~A.~A., et al.\ 2013, \mnras, 432, 2696
\bibitem[McQuinn et al.(2009)]{mcq09} McQuinn, M., Lidz, A., 
Zaldarriaga, M., et al.\ 2009, \apj, 694, 842
\bibitem[Oppenheimer et al.(2009)]{opp09} Oppenheimer, B.~D.,
Dav{\'e}, R., \& Finlator, K.\ 2009, \mnras, 396, 729 
\bibitem[Planck Collaboration(2015)]{pla15} Planck Collaboration 2015, submitted to \aap (arXiv:1502.01589)
\bibitem[Schaye(2006)]{sch06} Schaye, J.\ 2006, \apj, 643, 59 
\bibitem[Shapiro 
\& Giroux(1987)]{sha87} Shapiro, P.~R., \& Giroux, M.~L.\ 1987, \apjl, 321, L107
\bibitem[Springel(2005)]{spr05} Springel, V.\ 2005, \mnras, 364, 1105
\bibitem[Robertson et al.(2015)]{rob15} Robertson, B.~E., 
Ellis, R.~S., Furlanetto, S.~R., \& Dunlop, J.~S.\ 2015, \apjl, 802, L19
\bibitem[Schaye(2001)]{sch01} Schaye, J.\ 2001, \apj, 559, 507 
\bibitem[Siana et al.(2015)]{sia15} Siana, B., Shapley, 
A.~E., Kulas, K.~R., et al.\ 2015, \apj, 804, 17 
\bibitem[Theuns et al.(1998)]{the98} Theuns, T., Leonard, A., 
Efstathiou, G., Pearce, F.~R., \& Thomas, P.~A.\ 1998, \mnras, 301, 478 
\bibitem[Tittley 
\& Meiksin(2007)]{tit07} Tittley, E.~R., \& Meiksin, A.\ 2007, \mnras, 380, 1369
\bibitem[Vanzella et al.(2016)]{van16} Vanzella, E., de Barros, S., Vasei, K., et al.\ 2016, arXiv:1602.00688 
\bibitem[Verner et al.(1996)]{ver96} Verner, D.~A., Ferland, 
G.~J., Korista, K.~T., \& Yakovlev, D.~G.\ 1996, \apj, 465, 487
\bibitem[Weisz et al.(2014)]{wei14} Weisz, D.~R., Johnson, 
B.~D., \& Conroy, C.\ 2014, \apjl, 794, L3
\bibitem[Wise et al.(2014)]{wis14} Wise, J.~H., Demchenko, 
V.~G., Halicek, M.~T., et al.\ 2014, \mnras, 442, 2560
\bibitem[Wong et al.(2008)]{won08} Wong, W.~Y., Moss, A., 
\& Scott, D.\ 2008, \mnras, 386, 1023 
\bibitem[Worseck et al.(2011)]{wor11} Worseck, G., Prochaska, 
J.~X., McQuinn, M., et al.\ 2011, \apjl, 733, L24 
\bibitem[Worseck et al.(2014)]{wor14} Worseck, G., Prochaska, 
J.~X., Hennawi, J.~F., \& McQuinn, M.\ 2014, arXiv:1405.7405 
\bibitem[Zackrisson et al.(2011)]{zac11} Zackrisson, E., 
Rydberg, C.-E., Schaerer, D., {\"O}stlin, G., 
\& Tuli, M.\ 2011, \apj, 740, 13 
\frenchspacing
\end{thebibliography}
\end{document}